\documentclass{pasj00}
\begin{document}
\SetRunningHead{Wen-Cong Chen}{On the origin of orbital period
change in WY Cancri}
\Received{2013 February 4}%{yyyy/mm/dd}
\Accepted{2013 April 5}%{yyyy/mm/dd}
%\Published{}%{yyyy/mm/dd}

\title{On the origin of orbital period change in WY Cancri: a genuine angular momentum loss?}

%%% begin:list of authors
% Do NOT capitalize all letters in "textsc".
\author{Wen-Cong Chen %
  }
\affil{School of Physics, Shangqiu Normal University, Shangqiu
476000, Henan, China} \email{chenwc@pku.edu.cn}

%%% end:list of authors

%%% Please use the following style in case that sorting by
%%% affiliation is impossible.
%
% \author{%
%   D-Firstname \textsc{D-Familyname}\altaffilmark{1}
%   E-Firstname \textsc{E-Familyname}\altaffilmark{1,2}
%   and
%   F-Firstname \textsc{F-Familyname}\altaffilmark{2}}
% \altaffiltext{1}{Address of Institute}
% \email{ddddd@xxx.xxx.xx.xx}
% \email{eeeee@xxx.xxx.xx.xx}
% \altaffiltext{2}{Address of Institute}

%% `\KeyWords{}' always has to be placed before `\maketitle'.
 %Do NOT move this preamble from here!

\maketitle

\begin{abstract}
WY Cancri is a short-period ($P$=0.829~d) eclipsing RS Canum
Venaticorum stars, and both components are late-type stars.
Recently, observations provided by photometric observations and
light time minima show that the orbital period of WY Cancri is
experiencing a secular decrease at a rate of  ${\rm d } P/{\rm d
}t=-1.2\times10^{-7}\rm d~yr^{-1}$. In this Letter, we attempt to
investigate if the period change of WY Cancri can originate from
the angular momentum loss. In calculation,  we assume that this
source has a high wind loss rate of $\sim10^{-10}~\rm
M_{\odot}~yr^{-1}$. To account for the observation, magnetic
braking demands a strong surface magnetic field of $\gtrsim 10000$
G like Ap/Bp stars. Furthermore, if this source may be surrounded
by a circumbinary disk, and 6\% of the wind loss feeds the disk,
tidal torque between the disk and the binary can offer the
observed angular momentum loss rate. Such a strong magnetic field
or an extremely high wind input fraction seem to be highly
unlikely.
\end{abstract}

\section{Introduction}

Angular momentum loss plays an important role during the evolution
of close binaries. For example, orbital angular momentum loss
mechanisms are the vital input physics in studying the evolution
of binaries by the stellar evolution code. In narrow binaries with
an orbital period of $P< 3~ \rm hr$, gravitational wave radiation
is a dominant angular momentum loss mechanism
(\cite{kraf62,faul71}). The evolution of relatively wide binaries
($P> 3~ \rm hr$) is mainly governed by magnetic braking, which is
driven by the coupling between the stellar winds and the magnetic
field (\cite{verb81}). The angular momentum loss rate by magnetic
braking is related with the mass, radius, magnetic field, and
angular momentum of the magnetically active star, and is a more
complicated issue. Nowadays, there exist several empirical
prescriptions such as the standard magnetic braking model
(\cite{rapp83}) and the reduced magnetic braking model
(\cite{sill00}), and so on. It is significant to test various
magnetic braking models by observational data.

Close binaries can be ideal probes in examining various magnetic
braking models. It is successful for magnetic braking theory in
explaining the period gap of cataclysmic variables (CVs). However,
the contamination of the light curve caused by the accretion
processes in CVs results in a tiny change rate of orbital periods
($\dot{P}\sim 10^{-5} - 10^{-4} \rm s\, yr^{-1}$). Therefore, non
mass-transferring pre-CVs such as NN Serpentis may be credible
candidates (\cite{brin06,chen09}). In this Letter, we attempt to
explore if the angular momentum loss can be responsibel for the
orbital period change of WY cancri, which originated from the
catalog of \citet{stra93}.

WY Cancri is a member of short-period eclipsing RS Canum
Venaticorum (RS CVn) stars, which defined by \citet{hall76}. As a
single-lined spectroscopic binary, the spectrum types of two
components are G5 V and approximately M2, respectively
(\cite{arev99}). Employing the optical photometry of WY Cancri,
the system parameters have been derived (the detailed parameters
see also Table 1, \cite{zeil90,heck98}). This source is a typical
detached binary, and orbital period $P= 0.829~\rm d$. By
photometric observations and light time minima, it is discovered
that the orbital period of this source is continuously decreasing
at a rate of $\dot{P}=-1.44\times10^{-8}\rm d~yr^{-1}$
(\cite{hall80}). Recently, by $O - C$ diagram analysis of WY
Cancri \citet{tian09} reported that the orbital period is
experiencing a secular decrease at a rate of
$\dot{P}=-1.2\times10^{-7}\rm d~yr^{-1}$. Based on the period
decrease rate caused by magnetic braking given by \citet{guin88},
\citet{tian09} obtained that the period change rate
$\dot{P}=-4.5\times10^{-9}\rm d~yr^{-1}$, which is very difficult
to account for the observation. Therefore, it is an interesting
work to explore the genuine origin of the secular period change of
WY Cancri.

\begin{longtable}{lll}
  \caption{System parameters of WY Cancri.}\label{tab:LTsample}
      \hline
      & \citet{zeil90}&\citet{heck98}\\
      \endfirsthead
      \hline
      \endhead
      \hline

      primary mass     & $0.82\rm~ M_{\odot}$ &$0.81\rm~ M_{\odot}$  \\
      primary radius   &$0.94\rm~ R_{\odot}$  & $0.93\rm~ R_{\odot}$  \\
      secondary mass   & $0.31\rm ~M_{\odot}$ &$0.31\rm ~M_{\odot}$  \\
      secondary radius &$0.59\rm~ R_{\odot}$  & $0.58\rm~ R_{\odot}$  \\
      mass ratio       &$0.356\pm0.176$       & $0.384\pm0.099$ \\
      \hline
     \end{longtable}

\section{Analysis for angular momentum loss}
\subsection{Observed value}
Considering a close binary with a circular orbit consisting of a
primary star (of mass $M_{1}$), and a secondary star (of mass
$M_{2}$), its orbital evolution is governed by
\begin{equation}
\frac{\dot{J}}{J}=\frac{\dot{P}}{3P}+\frac{\dot{M}_{1}}{M_{1}}+\frac{\dot{M}_{2}}{M_{2}}-\frac{\dot{M}_{\rm
T}}{3M_{\rm T}},
\end{equation}
where $J=2\pi a^{2}M_{1}M_{2}/(M_{\rm T}P)$ is the total orbital
angular momentum of the binary, $a=(GM_{\rm
T}P^{2}/4\pi^{2})^{1/3}$ is the binary separation, $G$ is the
gravitational constant, $M_{\rm T}$ is the total mass of the
binary.

The primary of WY Cancri is a solar-like star, which has a stellar
wind-loss rates of $10^{-14} - 10^{-10}~\rm M_{\odot}~yr^{-1}$,
which depend on the stellar rotation rate (see also
\cite{wood02}). The mass-loss rates of several active M dwarf
stars were estimated to be $\sim10^{-10}~\rm M_{\odot}~yr^{-1}$
(\cite{mull92}). However, subsequent works presented contradictive
results, which are 2 -5 orders of magnitude lower than the one
derived by \citet{mull92} (\cite{lim96,oord97,wood01,warg02}).
Recently, employing 3D magnetohydrodynamical numerical simulations
\citet{vido11} found that the mass-loss rates of M dwarf V374 Peg
are $4\times10^{-10}~\rm M_{\odot}~yr^{-1}$. Therefore,
$\dot{M}_{1}/M_{1}$, $\dot{M}_{2}/M_{2}$, and $\dot{M}_{\rm
T}/M_{\rm T}$ are at least 2 - 3 orders of magnitude smaller than
$\dot{P}/{P}=-1.45\times10^{-7}\rm yr^{-1}$. By the analysis
mentioned above, equation (1) yields
\begin{equation}
\dot{J}=\frac{\dot{P}}{3P}J.
\end{equation}
Adopting the observed data derived by \citet{heck98} and
\citet{tian09}, the angular momentum loss rate of WY Cancri is
\begin{equation}
\dot{J}=-4.3\times 10^{36} \rm g\,cm^{2}\,s^{-2}.
\end{equation}

\subsection{Magnetic braking}
Both mass exchange and angular momentum loss can be responsible
for the secular decrease in the orbital period of the binaries.
When the material is transferred from the more massive primary to
the less massive secondary, the orbital period continuously
decreases. However, WY Cancri is a typical detached binary that
has no mass exchange. Therefore, the angular momentum loss by
magnetic braking may lead to the period change of this source.

\subsubsection{Standard magnetic braking model}

Some works studied the solar wind loss and the rotational rates of
solar-type stars in open clusters, and yielded the standard
magnetic braking model (\cite{webe67,skum72,mest87}). The loss
rate of angular momentum given by \citet{rapp83} can be written as
\begin{equation}
\dot{J}_{\rm smb}\simeq -3.8\times
10^{-30}MR_{\odot}^4(R/R_{\odot})^{\gamma}\omega^3\,\quad \rm
g\,cm^{2}\,s^{-2},
\end{equation}
where $M, R, \omega$ are the mass, the radius, and the angular
velocity of the magnetically active stars, respectively; $\gamma=0
-4$ is a dimensionless parameter. For WY Cancri, when $\gamma=0$,
we can calculate the angular momentum loss rate as
\begin{equation}
\dot{J}_{\rm smb}\simeq -9.0\times 10^{34}\rm g\,cm^{2}\,s^{-2},
\end{equation}
which is 2 orders of magnitude smaller than the observed result
given by equation (3).

%\subsubsection{Reduced magnetic braking model}
%According to the standard magnetic braking model, the spin-down of
%rapidly rotating stars should be extremely fast (\cite{pins90}).
%However, low-mass stars with a rapid rotation were detected in
%young open clusters (\cite{stau87}). All of the direct and
%indirect observational hints in low-mass stars and open clusters
%implies that the standard magnetic braking model overestimates the
%angular momentum loss rate for rapid rotators with a spin-period
%less than 2.5 - 5 d (\cite{andr03}). Therefore, \citet{sill00}
%developed a reduced magnetic braking description to explain the
%existence of low-mass stars with a highly rotation rate. It is
%clear that this angular momentum loss rate is too smaller to
%account for the observed period change seen in WY Cancri.

\subsubsection{Analytical magnetic braking model}
To account for the formation of black-hole low-mass X-ray binaries
with a short orbital-period ($P<1~\rm d$), \citet{just06} proposed
an anomalous magnetic braking mechanism of Ap/Bp stars. In their
work, an analytical magnetic braking description is given by
\begin{equation}
\dot{J}_{\rm amb}=-\frac{2\pi}{ P}(GM)^{-1/4}B_{\rm
s}R^{13/4}\dot{M}_{\rm wind}^{1/2},
\end{equation}
where $B_{\rm s}, \dot{M}_{\rm wind}$ are the surface magnetic
field, and the stellar wind loss rates of the magnetically active
stars, respectively.

Inserting the observed parameters of WY Cancri, surface magnetic
field of 10000 G, and stellar wind loss rates of
$10^{-10}~M_{\odot}\rm yr^{-1}$, equation (6) yields

\[
\dot{J}_{\rm amb}=-2.96\times 10^{36} \left(\frac{0.829\rm
d}{P}\right)\left(\frac{0.81M_{\odot}}{M}\right)^{1/4}\left(\frac{B_{\rm
s}}{10000\rm G}\right)
\]
\begin{equation} \left(\frac{R}{0.93R_{\odot}}\right)^{13/4}\left(\frac{\dot{M}_{\rm
wind}}{10^{-10}M_{\odot}\rm yr^{-1}}\right)^{1/2}\rm
g\,cm^{2}\,s^{-2}.
\end{equation}
In Figure 1, we plot the calculated angular momentum loss rate as
a function of the primary's surface magnetic field. It is clearly
seen that, if the primary of WY Cancri has a strong magnetic field
of $\gtrsim 10000$ G like Ap/Bp stars (\cite{moss89,brai04}),
magnetic braking can interpret the period change of this source.

\begin{figure}
 \begin{center}
   \FigureFile(95mm,95mm){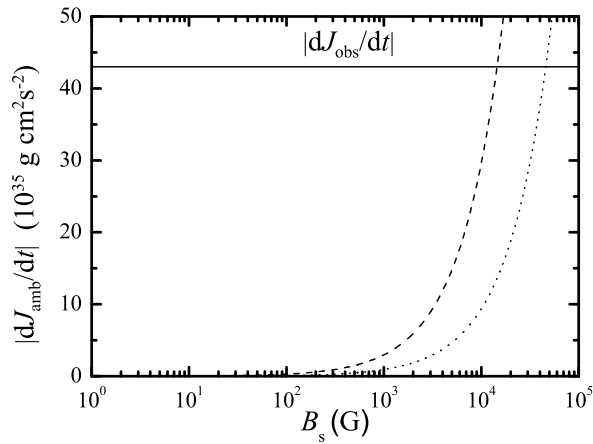}
    %%% \FigureFile(width,height){filename}
      \end{center}
  \caption{Angular momentum loss rate predicted by analytical magnetic braking model
  as a function of the primary's surface magnetic field. The dashed, and dotted curves correspond
  to the stellar wind loss rates of $10^{-10}$, and $10^{-11}~\rm M_{\odot}~yr^{-1}$,
  respectively. Horizontal solid line represents the inferred value in observation.}\label{fig:sample}
\end{figure}

\subsection{Circumbinary disk}
In fact, there may exist other efficient mechanism extracting
angular momentum from the binary such as circumbinary (CB) disk
(\cite{chen06}). CB disk may originate from a slow wind near the
orbital plane forming by the lost material during the mass
transferring of binary systems (\cite{heuv73,heuv94}), or the
remnant of common envelope (CE) that cannot be entirely ejected
during CE phase (\cite{spru01}). The tidal torques induced by the
gravitational interaction between the inner edge $r_{\rm i}$ of
the CB disk and the binaries can efficiently extract the orbital
angular momentum from the binary systems.

If WY Cancri is surrounded by a CB disk, and a fraction $\delta$
of the stellar wind from the primary feeds into the CB disk,
\citet{spru01} and \citet{taam01} deduced the following angular
momentum loss rate

\begin{equation}
\dot{J}_{\rm cb}=-\gamma\left(\frac{2\pi
a^2}{P}\right)\delta\dot{M}_{\rm wind}\left(\frac{t}{t_{\rm
vi}}\right)^{1/2},
\end{equation}
where $\gamma=\sqrt{r_{\rm i}/a}$ is a dimensionless parameter,
$t$ is the timescale of the stellar wind feeding the CB disk.
Assuming that the CB disk abides by the standard $\alpha$
prescription (\cite{shak73}), the viscous timescale at the inner
edge of the disk $ t_{\rm vi}=2\gamma^{3}P/(3\pi\alpha_{\rm
SS}\beta^{2}), $ where $\alpha_{\rm SS}$, and $\beta$ are the
viscosity parameter, and the dimensionless parameter described the
scale height at $r_{\rm i}$ of the CB disk, respectively.

For WY Cancri, the binary separation $a=2.68\times 10^{11}~\rm
cm=3.85~R_{\odot}$. Taking the typical parameters $\gamma=1.3$,
 $\alpha_{\rm SS}=0.01$, $\beta=0.03$ (\cite{chen06}), then the viscous timescale $
t_{\rm vi}=118~\rm yr$. We will adopt an extremely high stellar
wind loss rate of $\sim 10^{-10} M_{\odot}~\rm yr^{-1}$, and also
assume that phase during which the wind has been that extreme has
lasted for about 0.2 Gyr (see also \cite{wood02}). Using the above
parameters, equation (8) yields

\[
\dot{J}_{\rm cb}=-3.4\times 10^{36} \left(\frac{P}{0.829\rm
d}\right)^{-1}\left(\frac{a}{3.85R_{\odot}}\right)^{2}\left(\frac{\delta}{0.05}\right)
\]
\begin{equation} \left(\frac{\dot{M}_{\rm
wind}}{10^{-10}M_{\odot}\rm yr^{-1}}\right)\left(\frac{t}{0.2\rm
Gyr}\frac{118\rm yr}{t_{\rm vi}}\right)^{1/2}\rm
g\,cm^{2}\,s^{-2}.
\end{equation}
Figure 2 shows angular momentum loss rate extracted by a CB disk
as a function of the wind feeding fraction $\delta$. As shown in
this figure, for a high wind feeding fraction of 0.06
\footnote{This wind input fraction would be 1 - 2 orders of
magnitude higher than that adopted by Spruit \& Taam (2001).}, the
CB disk can be responsible for the orbit decay of WY Cancri.

\begin{figure}
 \begin{center}
   \FigureFile(95mm,95mm){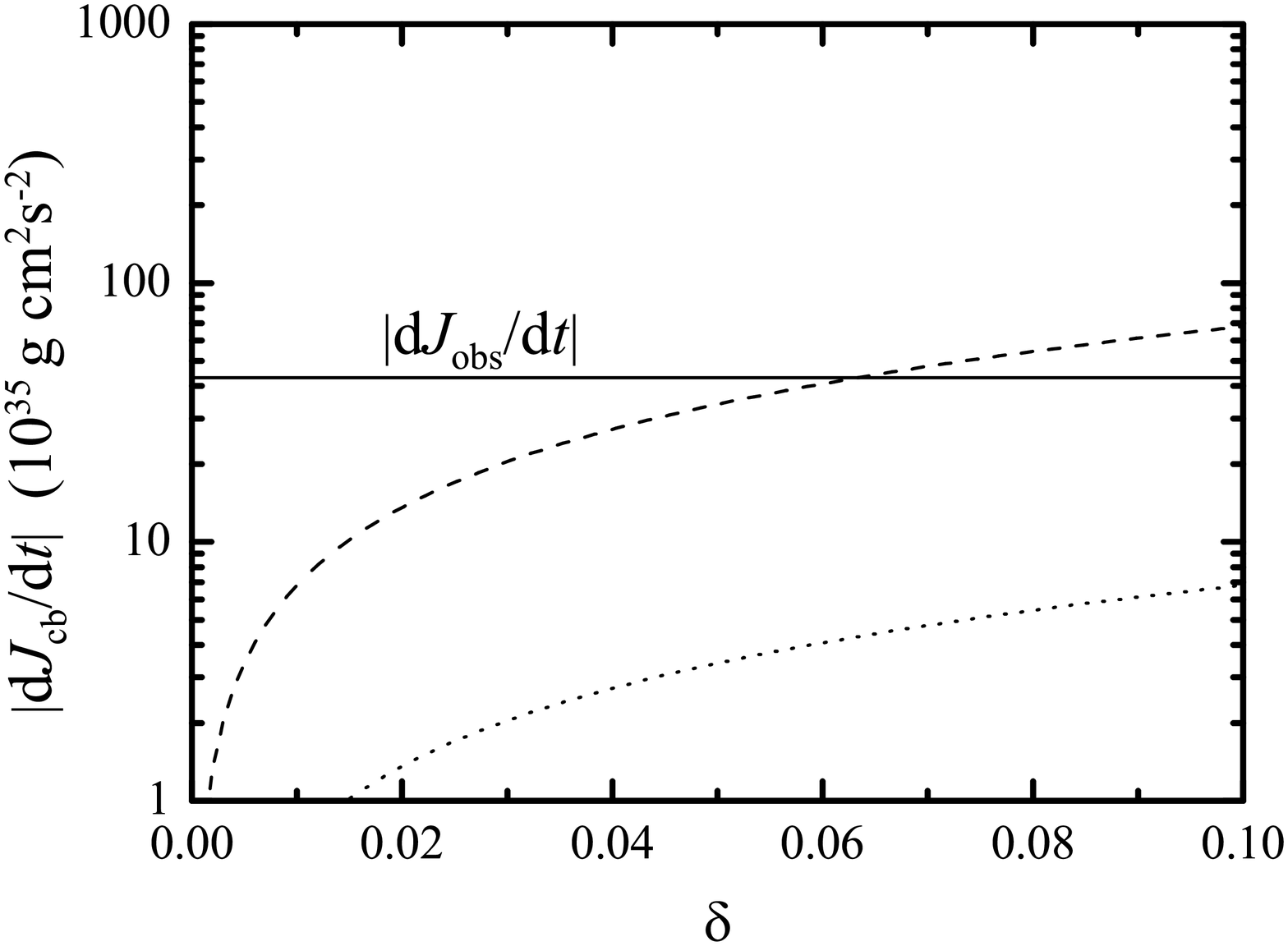}
    %%% \FigureFile(width,height){filename}
      \end{center}
  \caption{Angular momentum loss rate driven by a CB disk
  as a function of the wind feeding fraction $\delta$. Cases plotted
  by different curves are similar to Figure 1.}\label{fig:sample}
\end{figure}

\section{Summary and Discussion}

In this Letter, we have tested if the angular momentum loss can be
responsible for the secular period change of WY Cancri. Our main
results are summarized as follows.

1. The period change of WY Cancri cannot be explained by the
standard magnetic braking model, in which the angular momentum
loss rate is 2 orders of magnitude lower than the observed value.

2. If the primary of WY Cancri has a strong magnetic field of
$\gtrsim 10000$ G , and a high wind loss rate of $\sim10^{-10}~\rm
M_{\odot}~yr^{-1}$, magnetic braking description given by
analytical method can account for the observation. Although the
magnetic activity of RS CVn stars is drastic
(\cite{dona95,oste99,koch13}), such a high magnetic field is still
unfrequent. We expect that Zeeman Doppler imaging method in this
source can constrain the surface magnetic field, and confirm or
negate magnetic braking mechanism.

3. If this source is enclosed by a CB disk, and assuming that 6\%
of the stellar wind loss rate of $\sim10^{-10}~\rm
M_{\odot}~yr^{-1}$ feeds the disk, the corresponding loss rate of
angular momentum can interpret the observed value. The mass of CB
disk can be estimated to be $\delta\dot{M}_{\rm
wind}t\sim10^{-3}~\rm M_{\odot}$, which is significantly higher
than the inferred value ($\sim10^{-9}~\rm M_{\odot}$) given by
\citet{muno06}. Therefore, such an extremely high wind input
fraction seem to be highly unlikely.

If the period decrease recently observed in WY Cancri is indeed a
secular phenomenon, a genuine angular momentum loss should be
employed. In an allowable parameters range, magnetic braking or CB
disk may be responsible for the orbit decay of WY Cancri. Of
course, the observed period decrease may also be a short-term
phenomenon, and is only a stage of long-period oscillation
(\cite{tian09}). If so, Applegate's mechanism ({which also
requires a strong magnetic field of $\sim 1000$ G, \cite{appl92}),
or the presence of a third body in a long orbit around the binary
may be responsible for the period decrease. Therefore, long-term
observation in multiwaveband for WY Cancri should be continuously
in progress in the future.

%%%%%%%%%%%%%%%%%%%%%%%%%%%%%%%%%%%%%%%

%\begin{longtable}{lll}
%  \caption{Sample of ``longtable"}\label{tab:LTsample}
%  \hline
%  name & value1 & value2 \\
%\endfirsthead
%  \hline
%  name & value & value2  \\
%\endhead
%  \hline
%\endfoot
%  \hline
%\endlastfoot
%  \hline
%  aaaaa & bbbbb & ccccc \\
%  ...... & ..... & ..... \\
%  ...... & ..... & ..... \\
%  ...... & ..... & ..... \\
%  xxxxx & yyyyy & zzzzz \\
%\end{longtable}

\bigskip

Acknowledgement: We are grateful to the anonymous referee for
his/her helpful comments that lead to an improvement of this
manuscript. This work was partly supported by the National Science
Foundation of China (under grant number 11173018), Program for
Science \& Technology Innovation Talents in Universities of Henan
Province, and Innovation Scientists and Technicians Troop
Construction Projects of Henan Province, China.

%\appendix
%\section{Method of .....}

%\section{Approximation of ...}

%\section*{Complete data}

%%%
% See the manual for the detail.
%%%

\end{document}